\documentclass[12pt,preprint]{aastex}
\slugcomment{In press ApJ Letters, 1 August 2008}

\begin{document}

\title{Detection of additional members of the 2003 EL61 collisional family via
near-infrared spectroscopy}

\author{E.L. Schaller\altaffilmark{1}, 
M.E. Brown\altaffilmark{1}}
\altaffiltext{1}{Division of Geological and Planetary Sciences, California Institute
of Technology, Pasadena, CA 91125}
\email{schaller@caltech.edu}

\begin{abstract}

We have acquired near-infrared spectra of Kuiper belt objects 
2003 UZ117, 2005 CB79 and 2004 SB60 with NIRC on the Keck I Telescope.
These objects are dynamically close to the core 
of the 2003 EL61 collisional family and were suggested to
be potential fragments of this collision by \citet{darin}.
We find that the spectra of 2003 UZ117 and 2005 CB79 
both show the characteristic strong water ice absorption features seen exclusively on 
2003 EL61, its largest satellite, and the six other known collisional
fragments.   
In contrast, we find that the near infrared spectrum of 2004 SB60 is essentially
featureless with a fraction of water ice of less than 5\%.  
We discuss the implications of the discovery of these additional 
family members for understanding the formation and evolution of this collisional family
in the outer solar system.

\end{abstract}

\section{Introduction}

The only known collisional family in the Kuiper belt was detected because of the unique
 spectral properties the family members (Brown et al. 2007).  
The near-infrared spectra of all other
non-volatile rich Kuiper belt objects (KBOs)
 lie on a continuum between those whose spectra
contain 
moderate amounts of water ice absorptions to those 
whose spectra  are  essentially
featureless \citep{2007Natur.446..294B, kris08}.
In contrast, the near-infrared spectra of 2003 EL61,
its largest 
satellite, and five other small KBOs
resemble laboratory 
spectra of pure crystalline water ice \citep{1999ApJ...519L.101B,kris06,2007ApJ...655.1172T,2007A&A...466.1185M,2007Natur.446..294B, kris08}. 
Remarkably, these objects are also relatively clustered in orbital element space.
The largest object, 2003 EL61, had been previously suggested to have
experienced a massive collision that imparted its
fast (4-hour) rotation, stripped off most of its
icy mantle leaving it with a density close to rock (2.7 g/cc),
and formed its two satellites \citep{2006ApJ...639.1238R, 2006ApJ...639L..43B, 2008AJ....135.1749L}.
\citet{2007Natur.446..294B} concluded that the four extremely water ice rich
 KBOs and the large satellite of 2003 EL61 were in fact
fragments of the icy mantle of the proto-2003 EL61 that had been ejected
during a massive collision.

While collisional families in the asteroid belt can be
identified by their dynamical clustering alone, families in the Kuiper belt are much 
harder to identify because the 
collisional ejection velocities can be a significant fraction of an object's
orbital velocity.  Collisional fragments can therefore be spread out over a wide range of
orbital element space with the fastest ejected 
fragments having
significantly different orbits from the family core. 
The 2003 EL61
family members were all identified on the basis of their unique surface
properties: they are the only known objects
in the Kuiper belt with extremely pure water ice spectra and neutral
visible colors. 
It is only because of this unique surface signature
that detection of the 2003 EL61 family members was possible.

In order to determine if other known KBOs could be fragments of the 
2003 EL61 collision, \citet{darin} integrated the orbits of 131 high inclination
KBOs to  determine their proper orbital elements
and minimum ejection velocities away from the 2003 EL61 family 
core.  They determined the minimum ejection velocity 
a KBO must have had in order to reach its present orbit 
from the modeled location of the family forming impact.
In Figure 1 we show object H-magnitude vs. minimum ejection velocity 
(from \citet{darin}) for
the known KBOs closest to the family core.  For KBOs in known 
resonances, we plot the ejection velocity 
accounting for resonance diffusion \citep{darin}.
A large fraction of these objects
have never been observed to determine if they have IR spectral signatures
consistent with the 2003 EL61 family members.   It is important to note that
though all of the known 2003 EL61 family members have neutral visible
colors, visible color or visible spectroscopy alone \citep{PinAlons2008} is not sufficient for family
member identification. There are many KBOs with neutral visible colors
that do not also have strong water ice absorptions and are not members
of the 2003 EL61 collisional family.  
Therefore, while visible colors can be used
to rule out objects as potential family members, near infrared 
spectroscopy to determine water ice absorption 
depths is necessary for definitive family member identification.

Identifying additional
family members and characterizing the 
extent of the spread of fragments throughout the Kuiper belt
 may provide insight into the physics of this giant impact event
in the outer solar system.
In this letter we present near-infrared spectra 
obtained with the Keck I telescope of KBOs 2003 UZ117, 2005 CB79
and 2004 SB60, objects that are located 67, 97, and 221 m/s away from the 
modeled 2003 EL61 family core respectively (Fig 1).  
2003 UZ117 and 2005 CB79 were suggested
to be the most likely additional family candidates by \citet{darin}.

\section{Observations and Results}  

Observations of 2003 UZ117, 2005 CB79, and 2004 SB60
 were performed with the Near Infrared Camera (NIRC)
on the Keck I telescope.  
Objects were identified in the field 
by their motion relative to the background
stars.  Targets were moved to the center of a
0.525 arcsecond wide slit and were then observed with either
a J-through-H order sorting filter and a 120 lines per millimeter grism
 (2003 UZ117 and 2005 CB79) or
an H-through-K order sorting filter and a 150 lines per millimeter grism
(2004 SB60).
In all cases, exposure times for the individual spectra were 200 seconds.
Targets were dithered along the slit in a five position nod pattern
with 5 arcseconds between each dither.
During observations the telescope tracked at each target's predicted
rate of motion.

\subsection{2003 UZ117 \& 2005 CB79}
Low resolution spectral observations in the H--K (1.4-2.5 micron) region are ideal for 
determining the presence or absence of water ice due to two
broad water ice absorption features centered at 1.5 and 2.0 microns
(Figure 2).  \citet{kris08} observed
45 KBOs and centaurs in this wavelength range
and parameterized all spectra with a two component model consisting of crystalline water
ice covering a fraction of the surface, f, mixed linearly with dark continuum component
with slope m, covering (1-f) of the surface.
They solved for m and f in each spectrum 
by using a least squares minimization scheme using the Powell method.
Family members clearly stood out because of their high fractions
of water ice (f $>$ 0.8) in this parameterization 
compared with all other objects in the survey (f $<$ 0.5).

Low resolution spectral observations at H--K
with NIRC are only practical for objects with visual magnitudes
less than $\sim$21.
Spectral observations in the J--H region of the spectrum (1.1-1.6 microns) 
allow us probe the spectra of objects
(vmag$\sim$21.5) that would
be too faint to detect or obtain sufficient signal to noise
in a reasonable amount of time with H--K spectroscopy.
Fortunately, the water ice absorption feature centered at $\sim$1.5 microns
is perfectly suited for detection
with J--H spectroscopy.
We can therefore observe KBOs up to half a magnitude fainter than previous
observations in H--K \citep{kris08} allowing 
us to detect strong water ice absorptions and  
determine if these smaller objects are also members of the 
2003 EL61 collisonal family.

J--H spectra of two dynamically likely family members, 
2003 UZ117 and 2005 CB79 (visual magnitudes of 21.3 and 21.2 respectively)
were obtained.
2003 UZ117 was observed on 23 September 2007 UT for a total integration time
of 4000 seconds.
2005 CB79 was observed on 21 and 22 May 2007 UT for a total integration time of 
12200 seconds.  Nearby G-type main sequence stars were observed 
within 0.1 airmass of the targets for telluric calibration.
Variable seeing, sky conditions, and occasionally poor tracking (especially on 
the 2005 CB79 observations) caused the individual 
spectra of each of the KBOs to be of varying quality, thus only the highest signal-to-noise
spectra were included and the rest discarded for this analysis. The final 
spectra consist of total integration times of 2800 seconds for 2003 UZ117 and 5200 seconds for 2005 CB79.  
Spectral data reduction was carried out using standard procedures as described
in Brown (2000) and \citet{kris08}.

The final 2003 UZ117 and 2005 CB79 spectra were
each binned over four pixels to increase the 
signal to noise per pixel.  We used a Gaussian weighting function
with a FWHM of 8 pixels resulting oversampled spectra with a spectral
resolution of $\sim$33.
In Figure 2 we show these spectra with one sigma error bars along with a model of 
pure water ice with 50 micron grain sizes at 50 K \citep{1998JGR...10325809G}.
We can quantitatively parameterize the amount of 
water ice using the same spectral model
as \citet{kris08} but considering the wavelength region from
1.1-1.6 microns.
Using a Powell minimization scheme 
to solve for the best fit of the parameters,
m and f, we find that the spectra of 
2003 UZ117 and 2005 CB79 are 
fit by water ice fractions of 0.92 $\pm{0.16}$ and 1.02 $\pm{0.10}$ respectively.
The high fraction of water ice given by this parameterization for 2005 CB79 
may be due to the fact that the water ice grain sizes on this object
are somewhat larger 
than those modeled 
by \citet{kris08}.
Nonetheless, the deep absorptions detected from $\sim$1.4-1.6 microns
on 2003 UZ117 and 2005 CB79 are clearly consistent with a 
high fraction of water ice (in contrast with all other known 
KBOs) but comparable to the other 2003 EL61 family members \citep{kris08}.

\subsection{2004 SB60}
2004 SB60, with a visual magnitude of 20.8 was
bright enough to be observed in 
H - K.  We observed it on 23 September 2007 
for a total integration time of 3000 seconds.
Data reduction was carried out as described above.
A nearby G-type main sequence star was observed
at similar (within .03) airmass of the
target.  Division of the 2004 SB60 spectrum by the star spectrum allowed us to calibrate
telluric features.  The resulting spectrum has a resolution r$\sim$160.

Using the same spectral parameterization of \citet{kris08} and
described above,
we find that the fraction of 
water ice in the spectrum of 2004 SB60
is consistent with 0\% and can be no more than 0.05 (Figure 2).
The low water ice fraction seen in the spectrum of 2004 SB60
is similar to that seen in many small
KBOs (see \citet{kris08}).  The spectrum of 2004 SB60 is clearly not consistent with the other
members of the 2003 EL61 collisional family that have water
ice fractions greater than 0.8 (Figure 2).

\section{Discussion}

2003 UZ117 and 2005 CB79 contain high fractions
of pure water ice on their surfaces and appear
dynamically related to the other members of the 2003 EL61
collisional family.  We therefore conclude that
they are members of this collisional family.
Thus far, all objects within
130 m/s of the modeled collision center \citep{darin}
have been shown to have the 
same unique strong water ice spectral
signatures.
2004 SB60, at 221 m/s from the family core,
has an essentially  featureless spectrum
inconsistent with 2003 EL61 and its family members 
but comparable to many other small KBOs. 
The flat spectrum we observed is consistent with infrared 
photometric observations of 2004 SB60
with HST by  \citet{2007DPS....39.3906S}.
Infrared spectral or spectrophotometric observations of more potential fragments with minimum
ejection velocities from $\sim$150 to $\sim$300 m/s
could help constrain the extent to which fragments
were scattered throughout the Kuiper belt
and could tell us about the physics of the giant impact itself.
In addition to the distribution of fragments,
any collisional model would also need to explain the the
presence of the two satellites of 2003 EL61.

Though it does not have the spectral signature of the fragments
of the 2003 EL61 collision,
2004 SB60 is an interesting object in its own right.
\citet{2008Icar..194..758N} have found this object
to be one of the few high inclination small binary objects
in the Kuiper belt.
 It is worth noting 
the possibility that 
not all fragments of the 2003 EL61 collision
necessarily have the unique strong water ice signature.  
Fragments from different locations in the initial 
parent body may have had different initial compositions.
However, without
this unique spectral signature, identification of an object as a family
member is currently impossible. 

Another object near the center of the 2003 EL61 
family is the third largest KBO, 2005 FY9.  2005 FY9 has a velocity closer to the
center of the collision (150 m/s) than one of the known fragments
but has a spectrum dominated by methane, not water ice absorptions.
We find it intriguing that two of the largest KBOs are so close to each other in 
orbital element space but can think of no reason why they should be 
physically related.

The spectral signature of extremely pure water ice found 
on 2003 EL61 and its fragments and nowhere else in the Kuiper belt
is the only reason definitive identification
of the collisonal family was possible.   
A collision with a non differentiated parent body would likely
not produce such a distinct spectral signature in its fragments.
Future large scale surveys of the Kuiper belt such as PanSTARRS and
LSST are expected to increase
the numbers of known KBOs by over an order of magnitude.  With higher
numbers of objects, overdensities
in certain regions may be revealed and families without unique spectral signatures
may be able to be identified by their dynamics alone.

% {\it Acknowledgments:} 
\acknowledgments
The data presented herein were obtained at the W.M.
Keck Observatory, which is operated as a scientific partnership among the California Institute of Technology, The University of California and the National 
Aeronautics and Space Administration.  The observatory was made
possible by the generous financial support of the W.M. Keck Foundation.

\clearpage

\begin{figure}
\plotone{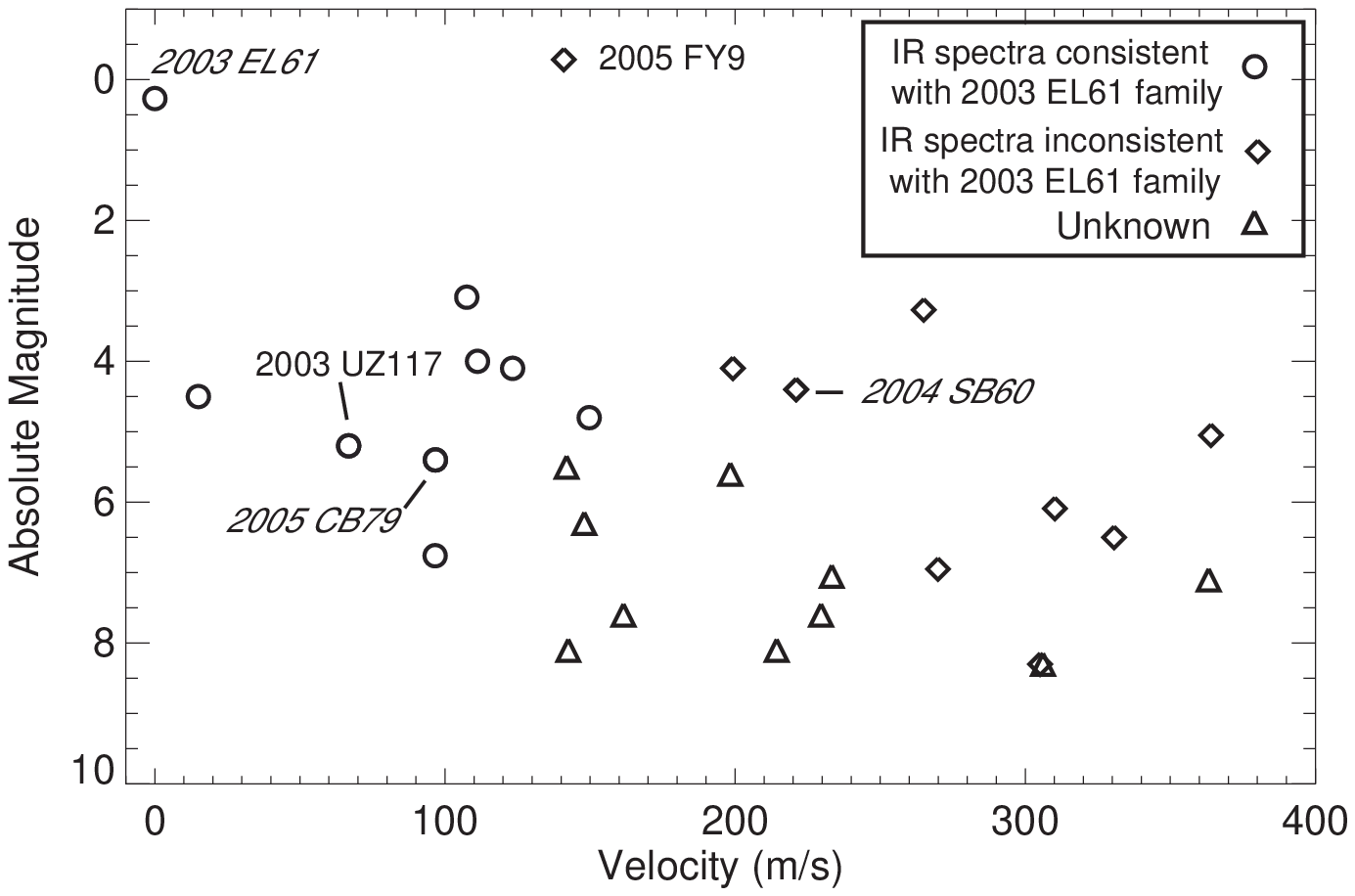}
\caption{Kuiper belt object absolute magnitude (H) vs. the minimum velocity required for a given KBO 
to reach its orbit from the modeled 2003 EL61 family forming collision 
(from Ragozzine \& Brown 2007 Tables 1 \& 2). 
Including 2003 UZ117 and 2005 CB79 (this paper), all of the objects within $\sim$130 m/s
have similar infrared spectra 
showing extremely deep absorptions due to water ice on their surfaces.  
These are the only
KBOs with such strong absorption features.
2004 SB60 was found to have an featureless infrared spectrum inconsistent
with the other 2003 EL61 family members.}
\end{figure} 

\clearpage

\begin{figure}
\epsscale{0.6}
\plotone{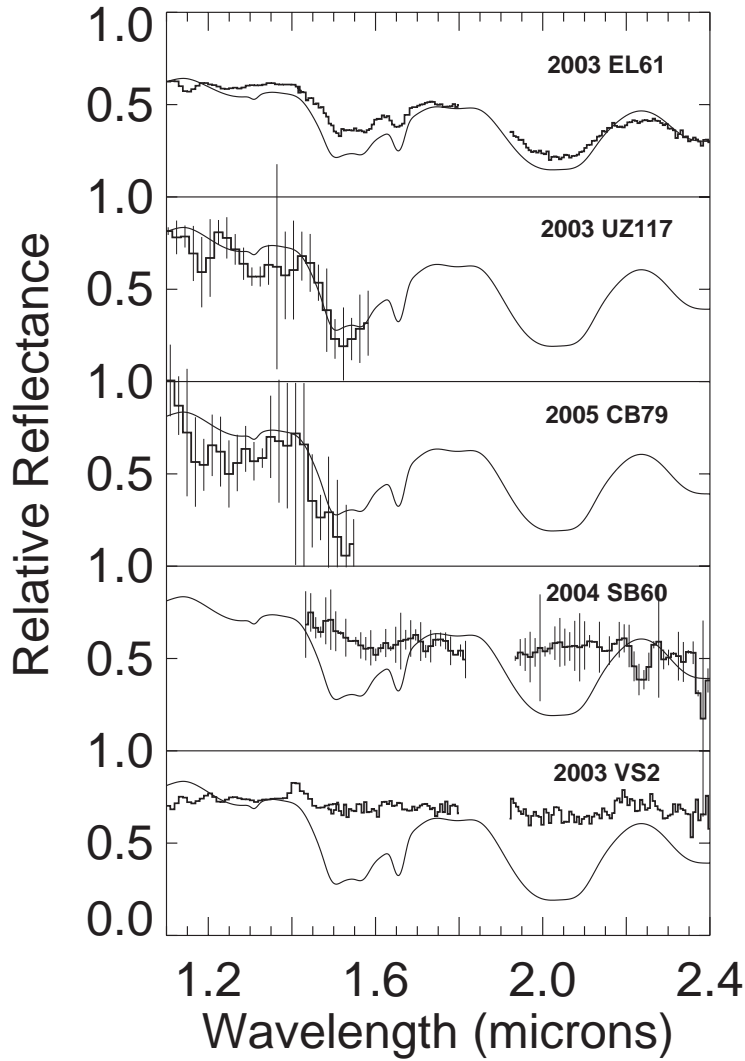}
\caption{Near infrared spectra of 2003 EL61 and its collisional family
members show strong signatures of nearly pure water ice on their surfaces.
Shown are the spectra of five KBOs along with a pure 
crystalline water ice model \citep{1998JGR...10325809G}.
We observed KBOs 2003 UZ117, 2005 CB79 and 2004 SB60, objects 67, 97 and 221 m/s
away from the modeled 2003 EL61 family core \citep{darin}
to determine if their near infrared spectra contained the characteristic 
strong water ice absorptions.
Spectra  of 2003 UZ117
and 2005 CB79 show strong absorptions at 1.4-1.6
microns consistent with the presence of high fractions
of water ice on their surfaces. In contrast, 
the spectrum of 2004 SB60 is essentially featureless.  The
flat spectrum of 2003 VS2 (from \citet{kris08}), a typical non-family
member KBO, is shown for comparison.}
\end{figure}

\end{document}